# An Efficient Anonymous Authentication Scheme Using Registration List in VANETs

Alireza Aghabagherloo
Electrical Engineering Department
Sharif University of Technology
Tehran, Iran
aghabagherlo.alireza@ee.sharif.edu

Javad Mohajeri
Electronics Research Institute
Sharif University of Technology
Tehran, Iran
mohajer@sharif.edu

Mahmoud Salmasizadeh
Electronics Research Institute
Sharif University of Technology
Tehran, Iran
salmasi@sharif.edu

Mahmood Mohassel Feghhi
Faculty of Electrical and Computer Engineering
University of Tabriz
Tabriz, Iran
mohasselfeghhi@tabrizu.ac.ir

*Abstract*—Nowadays, Vehicular Ad hoc Networks (VANETs) are popularly known as they can reduce traffic and road accidents. These networks need several security requirements, such as anonymity, data authentication, confidentiality, traceability and cancellation of offending users, unlinkability, integrity, undeniability and access control. Authentication is one of the most important security requirements in these networks. So many authentication schemes have been proposed up to now. One of the well-known techniques to provide users authentication in these networks is the authentication based on the smartcard (ASC). In this paper, we propose an ASC scheme that not only provides necessary security requirements such as anonymity, traceability and unlinkability in the VANETs but also is more efficient than the other schemes in the literatures.

## I. Introduction

Many road accidents can occur because of the ignorance of the driver about road conditions [1]. VANETs are applied for communication between vehicles and Road side Units (RSUs) and improving road safety. These networks consist of Trusted Authority (TA), Road Side Unit (RSU), and On-Board Unit (OBU). There are V2V and V2R communications in these networks, and according to [2], there is a secure channel between the RSUs and the TA. These communications are shown in Fig. 1. In this figure, suppose that the OBU1 and the OBU2 are moving at high speeds, given that these cars are not in sync with each other, an accident is more likely to happen. The VANET can prevent this accident. In these networks, each OBU acts as a node and sends information to the nearest OBU. In these networks, important security requirements are authentication, confidentiality, privacy and anonymity, traceability and cancellation of offending users, integrity, authenticity, undeniability, access control, public verifiability and unlinkability.

Authentication of the sender and data are two important security requirements in VANETs, because the source of the transmitted data should be verified to make sure the validity of the received data. Proposed authentication schemes for these networks can be divided into 4 categories. 1) Symmetric encryption-based authentication schemes [3-4] that do not meet the requirement of public verifiability and undeniability, 2) Asymmetric encryption-based authentication schemes that are divided into PKI and ID-based schemes [5-6],

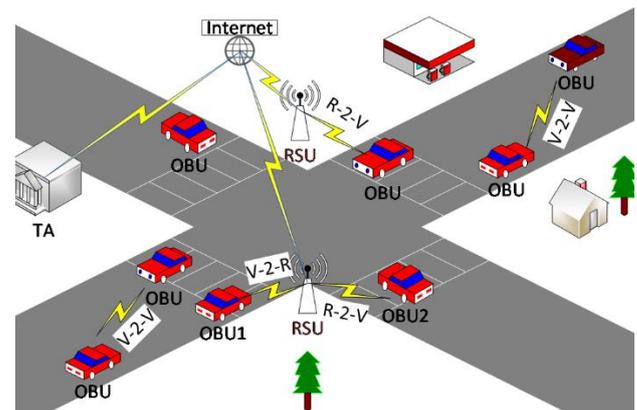

Fig. 1. Elements of vehicular ad-hoc networks.

3) Physically Unclonable Function (PUF) based schemes [7] and 4) Blockchain-based schemes [8]. Among these schemes, ID-based schemes are more efficient and secure. One of the well-known ID-based schemes is the Conditional Privacy-Preserving Authentication (CPPA) [9-10] that provides necessary security requirements in VANETs. However, if the infiltration to one of the TPD occurs, this scheme cannot provide several security requirements such as privacy-preserving and unlinkability.

Other well-known ID-based schemes are the Authentication based on Smart Card (ASC) schemes [2, 11,12]. Authors in [2] and [12] proposed ASC schemes that provide many security requirements in VANETs; however, they don't provide unlinkability in a RSU's range, and the TA cannot prevent offending IDs registration in the registration phase. Also, the presented scheme in [2] doesn't provide the pass change phase. Authors in [11] proposed another scheme that can prevent offending IDs because of using the registration list, which is not an efficient scheme.

In this paper, a regular ASC scheme will be proposed that not only provides security requirements such as anonymity, traceability and unlinkability, but also is more efficient than other ASC related works.

The paper is organized as follows: we review typical ASC schemes in Section II. Then in Section III, we explain the weakness of the ASC schemes. Section IV presents our proposed scheme, and section V evaluates the security and performance of the proposed scheme. Finally, we conclude the paper in Section VI.

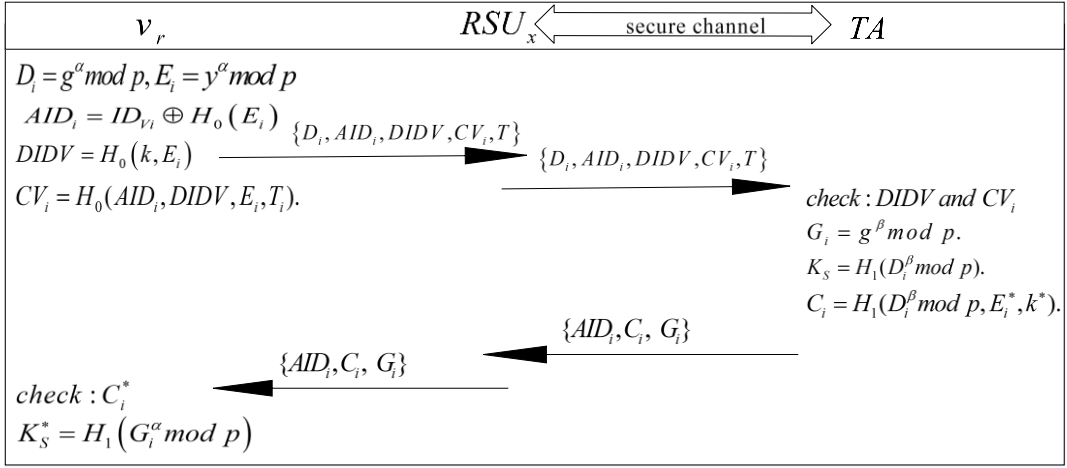

Fig. 2. User authentication phase of [2].

## II. REVIEW OF TYPICAL ASC SCHEMES IN VANETs

Typical ASC schemes are ID-based, which support V-2-V and V-2-R communications. These schemes, consists of three parts as follows:
- ✓ Trusted Authority (TA): This is the honest center of the network. TA produces primary parameters stored in OBUs and smart cards. Also, it participates in authenticates of OBUs entering at a RSU's range, and establish a session key for authenticated OBUs.
- ✓ Road Side Unit (RSU): RSU authenticates OBUs, which enter in its range with the help of TA, communicates wirelessly with the OBUs, and transfers the session keys determined by TA. It is assumed that there is a secure channel between RSUs and TA [2].
- ✓ On-Board Unit (OBU): Each vehicle is equipped with an OBU for secure communication with the around vehicles using a session key.

These schemes usually consist of five phases: **the user registration phase** where users register with the *TA*; **the user login phase** where vehicle owner with valid smart card logs in to a vehicle; **the user authentication phase** where vehicle efforts to establish a secure connection with the TA; **the data authentication phase** where authenticated datagrams are sent; **the password change phase** where the vehicular owner efforts to change his/her passwords [2].

Due to the desirable properties of [2] in the following, we review the proposed scheme in [2]. This protocol focused on the user registration phase, the user login phase and the user authentication phase.

In [2] it has been assumed that each vehicular user $V_i$ register with the *TA* once and the registration phase done through a secure channel without any adversary interference. In addition, an adversary will not be allowed to register with the *TA* using one or more identities.

The TA determines the system parameters as follows: $\mathbf{F}_p$ as a finite field of prime order p and generator g, a uniformly random element $x$ of $\mathbf{Z}_p$ and $y=g^x \bmod p$ as the master private and public keys, Also let s be the server's secret and $H_0$ and $H_1$ are two secure one-way hash functions.

Vehicular user $V_i$ picks its unique identifier $ID_{Vi}$ and transmits it to the *TA*. After getting the registration request, the *TA* first checks that the identifier has not been used before. Then it calculates $k = H_0(ID_{Vi} \oplus s)$ and delivers a smart card containing $\{ID_{Vi} ||k|| g|| p|| y|| H_0||H_1\}$ to vehicle owner $V_i$. To provide user login phase $V_i$ inserts the smart card in its vehicle. Throughout this procedure, the communication is supposed authentic.

Whenever $V_i$ desires to establish a secure communication with the TA through the nearest RSU, it runs the subsequent steps to get a session key. You can see the details in Fig. 2.

Step 1: The user $V_i$ inserts its smart card in the vehicle's card reader and after getting back $k$ and $y$ selects a uniformly random integer α, and the timestamp T, then calculates the subsequent expressions:

$$D_i = g^\alpha \bmod p. \quad (1)$$

$$E_i = y^\alpha \bmod p. \quad (2)$$

$$AID_i = ID_{Vi} \oplus H_0(E_i). \quad (3)$$

$$DIDV = H_0(k, E_i). \quad (4)$$

$$CV_i = H_0(AID_i, DIDV, E_i, T_i). \quad (5)$$

At the end, $V_i$ forwards $\{D_i, AID_i, DIDV, CV_i, T\}$ to the RSU.

Step 2: Whenever the RSU gets a login request message from a car, it checks the timestamp. If it is valid, it concatenates its identity to the message and sends it to the TA.

Step 3: TA first checks the timestamp, if it was valid, it attempts to verify the received message through the subsequent relations:

$$E_i^* = (D_i)^x \bmod p. \quad (6)$$

$$k^* = H_0(AID_i \oplus H_0(E_i^*) \oplus s). \quad (7)$$

$$CV^*_i = H_0(AID_i, DIDV, E_i^*, T_i). \quad (8)$$

The request would be rejected, if $DIDV \neq H_0(k^*||E_i^*)$ or $CV_i^* \neq CV_i$. If not, TA accepts this request and produce a random number β, and calculates the following values:

$$G_i = g^\beta \bmod p. \quad (9)$$

$$K_S = H_1(D_i^\beta \bmod p). \quad (10)$$

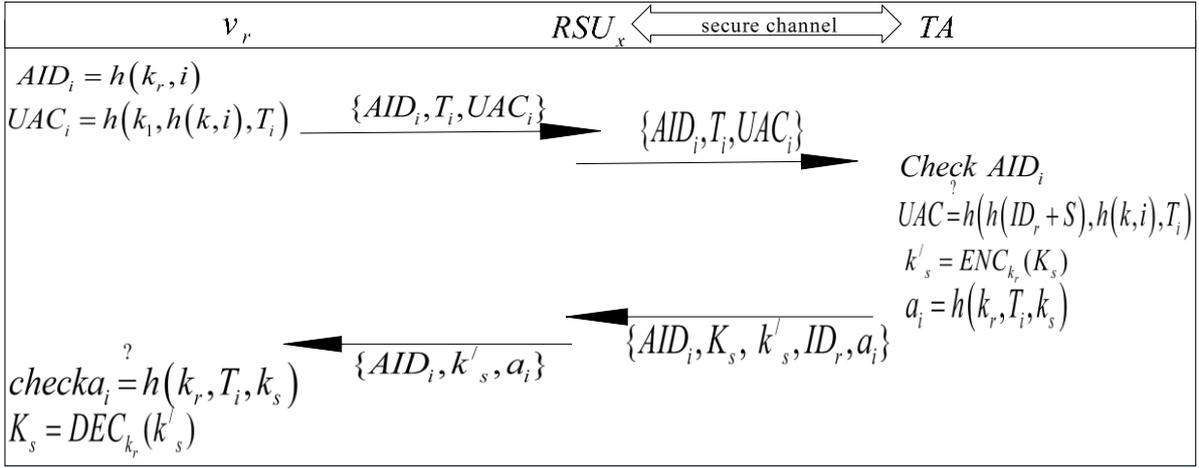

Fig. 3. User authentication phase of the proposed protocol.

$$C_i = H_1(D_i^\beta \mod p, E_i^*, k^*). \qquad (11)$$

Then TA forwards {$AID_i$, $C_i$, $G_i$} to the nearest RSU of the same vehicle through a secure channel. Then RSU broadcast this message.

Step 4: If a broadcasted message contains $AID_i$ then $OBU_i$ computes the following expressions:

$$C_i^* = H_1(G_i^\alpha \mod p, E_i, k). \qquad (12)$$

$$K_S^* = H_1\left(G_i^\alpha \mod p\right) \qquad (13)$$

If $C_i^* = C_i$ then it sets the $K_S^*$ as the session key.

### III. ANALYSIS OF THE ASC PROTOCOLS

In the previous section, we have review one of the well-known ASC schemes presented in [2]. It satisfies many security requirements such as authentication, confidentiality, privacy and anonymity, traceability, integrity, authenticity and undeniability and according to figure 4 this scheme is efficient among ASC schemes. Also, this scheme has been focused on the user registration phase, the user login phase and the user authentication phase and to provide other phases of ASC schemes authors of [2] referred to [12]. Authors in [12] proposed a hash chain based protocol for data authentication phase that not provides unlinkability in one RSU's range. Therefore, the proposed scheme in [2] cannot supply the unlinkability property for messages that transmit in one RSU's range. Also, however, [2] mentioned that the password changing phase is one of the ASC scheme phases but authors of [2] don't use PW and therefore, in [2] there is no solution to change the password. Another remark is that in [2] and [12], due to not employing a registration list, TA cannot prevent offending IDs registered in the registration phase but, in [11], TA can delete offending IDs from its registration list and don't let them be authenticated in the user authentication phase.

According to the outlined analysis, we want to propose a new ASC scheme that not only provides mentioned necessary security requirements such as authentication, confidentiality, privacy and anonymity, traceability, integrity, authenticity and undeniability, unlinkability and preventing offender users in the VANETs but also is more efficient than the other ASC schemes in the literatures.

### IV. THE PROPOSED SCHEME

The proposed scheme is an ID-based scheme using the registration list. The architecture of this scheme is similar to the scheme proposed in section 2. The phases are as follows:

- *User registration phase*

In this phase, a user with a valid ID registers with the TA. If we assume that h is a secure one-way hash function. In this phase, the TA with a server master key (S) chooses $ID_r$ and $PW_r$ for the r-th vehicle of the network, then it calculates $k_r = h(ID_r+S)$, $Z_r = k_r + h(PW_r)$ and h(k,1) for each vehicle. Then, the TA saves $ID_r$ and $h(k_r,0)$, the $OBU_r$ saves $ID_r$, $K_r$, $PW_r$, and $Z_r$, and the smart card saves $ID_r$ and $PW_r$.

- *User login phase*

In this phase, the vehicle owner is authenticated by OBU. First the vehicle owner inputs his smart card in the $OBU_r$. Then $OBU_r$ computes $h(PW_r)$ and checks equation $Z_r = k_r + h(PW_r)$ to be established. If the values are not matched, the smart card returns a fail.

- *User Authentication Phase*

In this phase, the OBU authenticates with TA, and then TA sends a session key to the OBU. Also, according to the assumptions in [2], we assume that there is a secure channel between the TA and the RSUs. Also, after the first authentication, the TA saves $h(k_r,2)$ instead of $h(k_r,1)$. In the next session, we use $h(k_r,2)$ for the authentication process. In the same way, we use $h(k_r,i)$ for the i-th session.

Step1: OBU calculates $AID_i$ and $UAC_i$ as follows:

$$AID_i = h(k_r, i). \qquad (14)$$

$$UAC_i = h(k_r, h(k_r, i), T_i). \qquad (15)$$

then sends $\{AID_i, T_i, UAC_i\}$ to the nearest RSU.

Step2: Then, the RSU sends $\{AID_i, T_i, UAC_i\}$ to the TA via a secure channel. The TA checks whether $T_i$ is valid, and also checks if the $AID_i$ exists in its registration list. If so, then the TA calculates: $UAC^* = h(h(ID_r+S), h(k_r,i), T_i)$, and checks the equation $UAC = UAC^*$. If UAC is valid this message is authenticated, then the TA computes $k'_s$ and $a_i$ as follows:

$$k'_s = ENC_{k_r}(K_s) \quad (16)$$

$$a_i = h(k_r, T_i, k_s) \quad (17)$$

Then the TA sends $\{AID_i, K_s, k'_s, a_i, ID_r\}$ to the RSU via a secure channel.

Step3: RSU saves $\{ID, AID_i, k_s\}$ in its authentication list ($L_{auth}$) to use it in the data authentication phase, then it sends $\{AID_i, k'_s, a_i\}$ to the $OBU_r$.

Step4: $OBU_r$ first checks whether $a_i^* = h(k_r, T_i, k_s)$ exist if so $OBU_r$ can calculate $K_s = DEC_{k_r}(k'_s)$ (with symmetric encryption). The details of this phase are illustrated in Fig 3.

- *Data authentication phase*

We use the proposed idea of [11] for this phase Because of providing unlinkability requirement in RSU's range. If a vehicle in travel wants to issue traffic information, the OBU sends $\{T_2, m, \sigma_m\}$ to the RSU and to other vehicles, where $\sigma_m = h(T2, m, K_s)$.

RSU receives $\{T_2, m, \sigma_m\}$ then checks whether timestamp $T_2$ is the latest. Then, the RSU finds out the item $\{K_s\}$ in its authentication list ($L_{auth}$), which satisfies $\sigma_m = h(T_2, m, K_s)$. If such an item is found in $L_{auth}$, the massage is valid; otherwise, if the equation is not satisfied in $L_{auth}$, the message is invalid. After message authentication of RSU, the RSU saves the $\{T_2, m, ID_r\}$ in the message authentication list of RSU ($L_m$), then notification message is produced by the RSU, consisting of the bloom filter. This filter stores the hash value of the valid traffic messages and their timestamps [13]. This notification message is encrypted with the private key $SK_{RSU}$ of the RSU, which can prevent an attacker from modifying or forging the notification message.

After sending the notification massage by the RSU, the OBUs around this RSU receive this notification message. Then, this OBUs decrypt this message using the public key of the RSU. If one of these OBUs wants to verify the validity of the message $\{T_5, m, \sigma_m\}$ sent by the other OBUs, the OBU will compute $h(T_5, m)$ and check whether this value is in the notification message. If this value exists in the notification message, the received message is valid, otherwise, the OBU waits to the next notification message from the OBU.

- *Password changing phase*

In this phase, the vehicle owner that feels the password has been exposed, changes his password saved in his smart card. In this phase, the $OBU_r$ owner inserts his $ID_r$ and his old $PW_r$. The OBU verifies these imported values. If the values are matched, then the vehicle owner inserts his $ID_r$ and his new $PW_{rnew}$. The $OBU_r$ computes the new $Z_{rnew} = k_r + h(PW_{rnew})$ and saves new parameters in his memory.

## V. SECURITY AND EFFICIENCY ANALYSIS OF NEW SCHEME

### A. Security analysis and security proofs

We compare the security requirements of this proposed scheme, and also prove these security requirements:

Table. 1. Security comparison among the different schemes.

| Security requirement \ scheme | [2] | [11] | [12] | proposed scheme |
|---|---|---|---|---|
| ID preserving | + | + | + | + |
| Traceability | + | + | + | + |
| Unlinkability | - | + | - | + |
| Pass change phase | - | + | + | + |
| Prevent offender IDs | - | + | - | + |
| Efficiency | - | - | - | + |

**1) Identity privacy-preserving**: The OBUs only send the pseudo-identity (AID) when it reaches to the range of the RSU. Pseudo-identity is calculated by the equation $AID_i = h(k_r, i)$ and $UAC_i = h(k_r, h(k_r, i), T_i)$. According to the hash function properties, these values are random. Therefore, no polynomial-time malicious can gain the real identity of the OBU through the pseudo-identity AID. Hence, our proposed scheme has met the requirements of the identity privacy-preservation.

**2) Traceability and Prevention of offender IDs:** When the RSU encounters malicious messages, it can search the item $\{T_2, m, ID_r\}$ in its message list $L_m$, and then sends $\{ID_r\}$ to the TA. The TA can search the item $<ID_r>$ in its registration list. Then the TA can prevent this ID from sending a message by invalidating the offender IDs in the user authentication phase.

**3) Non-repudiation:** If the OBU wants to deny its message, the RSU can find the $\{T_2, m, ID_r\}$ according to the timestamp $T_2$ in the message list $L_m$ quickly, which contains the real identity of the OBU and its registration time. Thus, our scheme has met the requirements of non-repudiation, and the OBUs cannot deny their messages.

**4) Unlinkability:** According to the preimage resistant property of the hash function, $AID_i = h(k_r, i)$, $AID_{i+1} = h(k_r, i+1)$ and $UAC_i = h(k_r, h(k_r, i), T_i)$, $UAC_{i+1} = h(k_r, h(k_r, i+1), T_{i+1})$ are entirely different in the user authentication phase. Also, in the data authentication phase, the message format in the new proposed scheme is $\{T_2, m, \sigma_m\}$, where $\sigma m = h(T_2, m, k_s)$. Therefore, due to the properties of the hash function of two messages, the adversary cannot define if the two given messages are sent by the same $OBU_i$ using the message content, which achieves the unlinkability requirement.

**5) Modification of passwords**: A vehicle's owner can modify the passwords anywhere and anytime, whenever he/she considers the passwords are not secure, and its details are brought in password changing phase of proposed scheme.

These comparisons are illustrated in Table 1.

### B. Efficiency analysis and compare it with other schemes

ASC schemes consist of five phases, and each phase has computation costs and communication burden. Also, according to [14], the RSUs have power and bandwidth limitations. Also, they are close to each other, and we need a new authentication phase in each RSU's range. Therefore, low computation cost and communication burden in the user authentication phase is important. We define $T_e$=Modular exponentiation time, $T_m$=Multiplication of points on ECC time (secp256r1), $T_h$=Hash time (SHA-256), $T_{enc}$=symmetric encryption/decryption time and $T_\oplus$=XOR function time.

Table. 2. Efficiency comparison in the authentication phase of different schemes on each side of the VANET architecture.

| overHead | OBU | RSU | TA |
|---|---|---|---|
| [2] | $3T_e + 5T_h + T_\oplus = 14.43ms$ | 0 | $3T_e + 6T_h + 2T_\oplus = 14.46ms$ |
| [11] | $3T_m + 7T_h + 3T_\oplus = 60ms$ | $T_m + 4T_h + T_\oplus = 20.35ms$ | $2T_m + 9T_h + 5T_\oplus = 40.73ms$ |
| [12] | $T_e + 8T_h + T_{enc} + T_\oplus = 5.12ms$ | $T_h + T_\oplus = 0.03ms$ | $T_e + 8T_h + T_{enc} + 2T_\oplus = 5.12ms$ |
| proposed scheme | $3T_h + T_{enc} = 0.21ms$ | 0 | $4T_h + T_{enc} = 0.24ms$ |

In [2], the time of this operations using a smartphone (iPhone 6s) as the test platform and the specific parameters with system iOS 10.11, CPU Apple A9 C M9 coprocessor C up to 2.1 GHz, RAM 2GB, has been calculated as follows:

$T_e$=4.76ms, $T_m$=20.23ms, $T_h$=0.03ms, $T_{enc}$=0.12ms, and $T_\oplus$ is negligible.

Accordingly, Table 2 shows the computational complexity and implementation timing of the new proposed scheme and several ASC-based schemes in the user authentication phase. The results of Table 2 and figure 4 show low computational complexity and time of implementation of the new proposed scheme compared to the previous schemes.

Therefore, the new proposed scheme has a low computation and communication overhead compared to the other ASC schemes.

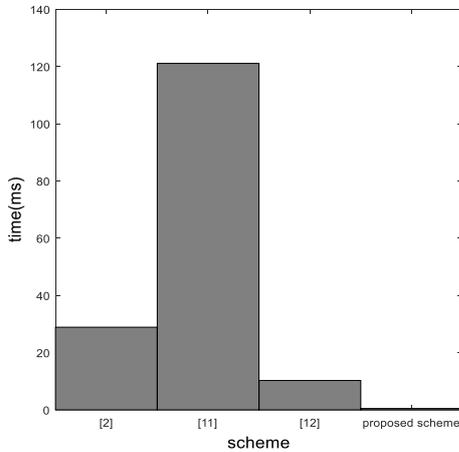

Figure. 4. Efficiency comparison in the authentication phase of different schemes.

## VI. CONCLUSION

In this paper, with regards to the vulnerability and inefficiency of ASC-based schemes, we have made some modifications to this schemes and improved the efficiency of the ASC schemes, while maintaining the desired features of the ASC schemes. Besides, the proposed scheme is resistant to many of the vulnerabilities of the previous schemes, furthermore according to figure 4 and table 2 the proposed scheme is more efficient than the similar works.

REFERENCES

[1] M. Kazemi, M. Delavar, J. Mohajeri and M. Salmasizadeh "On the Security of an Efficient Anonymous Authentication with Conditional Privacy-Preserving Scheme for Vehicular Ad Hoc Networks", Iranian Conference on Electrical Engineering (ICEE), pp. 510-514, 2018.

[2] C. M. Chen, B. Xiang, Y. Liu, and K. H. Wang, "A secure authentication protocol for Internet of vehicles," *IEEE Access*, vol. 7, pp. 12047-12057, 2019.

[3] X. Lin, X. Sun, X. Wang, C. Zhang, P. H. Ho, and X. Shen, "TSVC: Timed efficient and secure vehicular communications with privacy preserving," *IEEE Transactions on Wireless Communications*, vol. 7, no. 12, pp. 4987-4998, 2008.

[4] C. Zhang, X. Lin, R. Lu, P. H. Ho, and X. Shen, "An efficient message authentication scheme for vehicular communications," *IEEE Transactions on Vehicular Technology*, vol. 57, no. 6, pp. 3357-3368, 2008.

[5] A. Wasef, and X. Shen, "EMAP: Expedite message authentication protocol for vehicular ad hoc networks," *IEEE transactions on Mobile Computing*, vol. 12, no. 1, pp. 78-89, 2011.

[6] C. D. Jung, C. Sur, Y. Park, and K.H. Rhee, "A robust and efficient anonymous authentication protocol in VANETs," *Journal of Communications and Networks*, vol. 11, no. 6, pp. 607-614, 2009.

[7] M. Kim, W. Choi, A. Lee, and M.S. Jun, "PUF-based privacy protection method in VANET environment," *Advances in Computer Science and Ubiquitous Computing,* pp. 263-268, 2015.

[8] R. Shrestha, R. Bajracharya, A. P. Shrestha, and S. Y. Nam, "A new-type of blockchain for secure message exchange in VANET," *Digital Communications and Networks*, 2019.

[9] M. Bayat, M. Barmshory, M. Rahimi and M. R. Aref. "A secure authentication scheme for VANETs with batch verification." *Wireless Networks* vol. 21, pp. 1733-1743, 2015.

[10] D. He, S. Zeadally, B. Xu, and X. Huang, "An efficient identity-based conditional privacy-preserving authentication scheme for vehicular ad hoc networks," *IEEE Transactions on Information Forensics and Security*, vol. 10, no. 12, pp. 2681-2691, 2015.

[11] H. Zhong, B. Huang, J. Cui, Y. Xu, and L. Liu, "Conditional privacy-preserving authentication using registration list in vehicular ad hoc networks," *IEEE Access*, vol. 6, pp. 2241-2250, 2018.

[12] B. Ying, and A. Nayak, "Anonymous and lightweight authentication for secure vehicular networks," *IEEE Transactions on Vehicular Technology*, vol. 66, no. 12, pp. 10626-10636, 2017.

[13] T. W. Chim, S. M. Yiu, L. C. K. Hui, and V. O. K. Li, "SPECS: Secure and privacy enhancing communications schemes for VANETs," *Ad Hoc Netw.*, vol. 9, no. 2, pp. 189-203, 2011.

[14] "Technical Specification Dedicated Short Range Communications in Intelligent Transport Systems," Telecommunications Standards Advisory Committee (TSAC), Issue 1 Rev 1, 2017.